\documentclass[twocolumn,10pt]{IEEEtran}
\usepackage{bm,cite,float,amsmath,amssymb,amsthm}
\usepackage{multirow}
\usepackage{algorithm}
\usepackage[noend]{algpseudocode}
\usepackage{indentfirst}
\usepackage{xcolor}
\usepackage{makecell}
\usepackage{color}
\usepackage{bbding}
\usepackage{lipsum}
\usepackage{mathtools}
\usepackage{cuted}
\ifCLASSOPTIONcompsoc
  \usepackage[caption=false,font=normalsize,labelfont=sf,textfont=sf]{subfig}
\else
  \usepackage[caption=false,font=footnotesize]{subfig}
\fi

\makeatletter
\def\BState{\State\hskip-\ALG@thistlm}
\makeatother

\usepackage{amssymb}
\usepackage{amsmath}
\usepackage{graphicx}
\usepackage{cite}

\usepackage{balance}
\usepackage[utf8]{inputenc}

\usepackage{color,soul}

\usepackage{adjustbox}

\usepackage{verbatim}

\IEEEoverridecommandlockouts

\usepackage{graphicx,epstopdf}
\usepackage{epsfig}	
\usepackage{amsfonts,balance}
\usepackage{bbm}
\floatname{algorithm}{Algorithm}
\usepackage{lipsum}
\usepackage[colorlinks,urlcolor=blue,linkcolor=black,citecolor=blue,breaklinks=true]{hyperref}
\usepackage{url}
\newtheorem{theorem}{Theorem}

\newtheorem{corollary}{Corollary}

\makeatletter
\def\ScaleIfNeeded{%
	\ifdim\Gin@nat@width>\linewidth \linewidth \else \Gin@nat@width
	\fi } \makeatother

\begin{document}
	
	\title{Semantic Communications for  Vehicle-Based Mission-Critical Services: Challenges and  Solutions }
	\author{Hui~Zhou, Jiaying~Guo, Marios~Aristodemou, \\Zhaoyang~Du, Shen~Wang, Xiaolan~Liu, Soufiene~Djahel, and Celimuge Wu
		\thanks{Hui Zhou and Soufiene Djahel are with the Centre for Future Transport and Cities, Coventry University, U.K. (email:\href{mailto:hui.zhou@coventry.ac.uk}{hui.zhou@coventry.ac.uk}; \href{mailto:ae3095@coventry.ac.uk}{ae3095@coventry.ac.uk})}
		\thanks{Jiaying Guo and Shen Wang  are with the School of Computer Science, University College Dublin, Dublin, Ireland (email:\href{mailto:jiaying.guo@ucd.ie}{jiaying.guo@ucd.ie}; \href{mailto:shen.wang@ucd.ie}{shen.wang@ucd.ie})}
		\thanks{Marios~Aristodemou is with the Department of Computer Science, University of York, U.K. (email:\href{mailto:marios.aristodemou@york.ac.uk}{marios.aristodemou@york.ac.uk})}
        \thanks{Xiaolan~Liu  is with School of Electrical, Electronic and Mechanical Engineering, University of Bristol, U.K. (email:\href{mailto:xiaolan.liu@bristol.ac.uk}{xiaolan.liu@bristol.ac.uk}) (Corresponding author: Xiaolan Liu)} 
        \thanks{Zhaoyang Du and Celimuge Wu are with the Graduate School of Informatics and Engineering, The University of Electro-Communications, Japan (email:\href{mailto:duzhaoyang@uec.ac.jp}{duzhaoyang@uec.ac.jp}; \href{mailto:celimuge@uec.ac.jp}{celimuge@uec.ac.jp}) }
	}
\maketitle

%        \STRUC{brown means the structure of section}
\begin{abstract}
As mission-critical (MC) services such as Unmanned Aerial Vehicles (UAVs) based emergency communication and Internet of Vehicles (IoVs) enabled autonomous driving emerge, the traditional communication framework can not meet the growing demands for higher reliability and lower latency and the increasing transmission loads. Semantic Communication (SemCom), an emerging communication paradigm that shifts the focus from bit-level data to its context and intended task at the receiver (i.e., semantic level), is envisioned to be a key revolution in Sixth Generation (6G) networks. However, an explicit and systematic SemCom framework specifically tailored for Vehicle-based MC (VbMC) services has yet to be proposed, primarily due to the complexity and lack of analysis on their MC characteristics. In this article, we first present the key information-critical and infrastructure-critical vehicle-based services within the SemCom framework. We then analyze the unique characteristics of MC services and the corresponding challenges they present for SemCom. Building on this, we propose a novel SemCom framework designed to address the specific needs of MC services in vehicle systems, offering potential solutions to existing challenges. Finally, we present a case study on UAV-based rapid congestion relief, utilizing eXplainable AI (XAI) to validate the effectiveness of the proposed SemCom framework.
\end{abstract}


% \fi
%
% For peer review papers, this IEEEtran command inserts a page break and
\maketitle

\section{Introduction}
With the rapid growth of Unmanned Aerial Vehicle (UAV) and Internet of Vehicle (IoV) markets, various emerging Vehicle-based Mission-Critical (VbMC) services, including emergency communication and traffic control, have attracted significant attention from both academia and industry. Failure of these services may result in significant financial losses, safety hazards, or other critical consequences. Therefore, it is crucial to guarantee the stringent data transmission requirements to maintain the efficiency and reliability of VbMC services. However, transmitting large volumes of data, required by such services, with high reliability and low latency lays a significant burden on existing wireless networks, further highlighting the need for innovation solutions. 
 
Recently, semantic communications (SemCom) has been regarded as an emerging paradigm in the sixth-generation (6G) network to break through the traditional Shannon capacity limit,  aiming to transmit the important information related to the underlying communication task instead of the precise transmission of every bit or symbol of information \cite{weaver1949}. The SemCom paradigm offers new opportunities for the MC services by extracting the semantic information from service-related data (e.g., video streaming) and Control and Command (C\&C) signal, which reduces the volume of data that needs to be transmitted while maintaining the effectiveness of the services. 

Initial works on SemCom have focused on various transmitted data types, including text, voice, image, and video, where semantic information extraction and recovery methods, and metrics have been investigated \cite{Luo2022,hui2024}. In \cite{Luo2022}, the authors provided an overview of deep learning (DL) and end-to-end (E2E) communication based SemCom, with an emphasis on the text, voice, and image. In \cite{hui2024}, the authors have further extended the application of SemCom to emerging Extend Reality (XR) and Machine Learning (ML) algorithms by incorporating both semantic and effectiveness levels. However, most of the existing works mainly focused on a single data type without considering the complex environment and stringent requirements of VbMC services.

Several works in the literature have focused on securing SemCom against the increasing cyber threats, especially in MC services \cite{Du2023, Sag2023, Yang2024}. In \cite{Du2023}, the authors proposed novel security performance indicators for Internet-of-Things (IoT) under SemCom framework. While, \cite{Sag2023} identified multi-domain security vulnerabilities when using deep neural networks (DNNs) for SemCom, \cite{Yang2024} proposed secure SemCom techniques to tackle both information and semantic ML model threats. However, a comprehensive study of the unique challenges in applying SemCom to VbMC services, and potential solutions is still missing.

\begin{table*}[!htb]
    \caption{Implementation of vehicle-based mission-critical services under SemCom}
    \begin{center}
        \begin{tabular}{|c|c|c|c|c|c|}
            \hline
            \textbf{Criticality} &\textbf{Vehicle-based Services}  &\textbf{Task} &\textbf{Link}& \textbf{Data Modality}&\textbf{SemCom Information}\\
            \hline			
            \multirow{3}{*}{\makecell[l]{Information}}&\multirow{2}{*}{\makecell[l]{Emergency Communication}}&Map Construction&Backhaul&Video&Unique Local Area Information\\
            \cline{3-5}\cline{6-6}
            &&Evacuation&Downlink&Image&Evacuation Path\\
            %\hline
            %On-demand Communication &Live Streaming&Uplink&Video& Face\\
            \cline{2-6}
            &Rapid Congestion Relief &Traffic Light Control&D2D&Video&Queue Length \& Road Occupancy\\
            \hline
            \multirow{3}{*}{\makecell[l]{Infrastructure}}&\multirow{2}{*}{\makecell[l]{Search and Rescue}}  &\multirow{2}{*}{\makecell[l]{Civilians Tracking}}&Uplink&Video& Location of Person\\
            \cline{4-6}
             &&&D2D&C\&C&VoI \& AoI\\
            \cline{2-6}
            &Autonomous Driving &Real-time Decision &D2D&Video \& LiDAR point cloud& Situation-awareness Information \\
            \hline
        \end{tabular}
    \label{SemCom_summarization}
    \end{center}
\end{table*}

The main contributions of this article are:
\begin{itemize}
    \item Analyzing the typical VbMC services under the SemCom paradigm, including information-critical and infrastructure-critical services, where the corresponding link, data modality, and semantic information have been identified.
    \item Providing a concrete vision of the
unique characteristics and challenges of applying SemCom in VbMC services, with a particular focus on Knowledge Bases (KBs) synchronization, coupled multi-modal data processing, scalability, explainability, and security challenges.
    \item Proposing a novel SemCom framework to solve the above challenges, by integrating and exploiting swarm intelligence, large language models (LLMs), and explainable (XAI) techniques, to foster efficient cooperation among vehicles.
    \item Demonstrating the effectiveness of proposed SemCom framework through a case study, leveraging XAI-guided semantic feature extraction in UAV-based congestion relief scenarios. 
\end{itemize}

\section{Typical Vehicle-based Mission-Critical Services under SemCom Framework} % 

In this section, we analyze the typical VbMC services under the SemCom framework, and classify them into information-critical services and infrastructure-critical services, as summarized in Table.~\ref{SemCom_summarization}.

\subsection{Information-Critical Services}
Information-critical services focus on transmitting the required information reliably from the vehicle to the entity with a stringent latency requirement, such as in emergency and congestion relief scenarios.
\subsubsection{Emergency Communication}
Relying on the backhaul transmission via satellites or high-altitude platforms (HAPs), UAVs can be easily deployed as flying base stations (BSs) to provide emergency communication for user equipments (UEs). 
In emergency communication scenarios, UAVs can construct disaster area map and broadcast the emergency information (e.g., the evacuation path) to UEs. Empowered by the SemCom, UAVs can collaboratively extract unique semantic information from locally captured images for disaster area map construction leveraging a central server. Then, each UAV can customize a semantic evacuation path based on the position of each associated UE.

\subsubsection{Rapid Congestion Relief}
\label{subsec: congestion}

To achieve rapid traffic congestion relief, UAVs can provide high‐frequency, high-resolution monitoring data to local traffic light controllers (TLCs). In such scenarios, upon notification of an unexpected road closure, a UAV equipped with camera is dispatched to the most congested intersections affected by the diversion of a large number of vehicles \cite{Guo2025}. Instead of transmitting raw imagery, the UAV performs on-board semantic extraction, distilling key traffic features such as queue length and road occupancy. By processing semantic information at less than 100kbps compared to raw video streams exceeding 100 Mbps, the TLC can respond with update intervals in seconds, thus facilitating real-time inference for signal switching.

\subsection{Infrastructure-Critical Services}
The infrastructure-critical services focus on completing critical missions cooperatively within complex environments, such as in Search and Rescue (SAR) and autonomous driving scenarios.

\subsubsection{Search and Rescue}

UAV swarms can assist ground rescue teams to detect civilians and track their movements, thereby guiding responders towards those in need. Depending on the environmental constraints and mission requirements, the UAV fleet may need to dynamically and rapidly adapt its flight paths to achieve mission goals while maintaining reliable connectivity both within the swarm and with the rescue team. Therefore, ultra-reliable and ultra-low-latency communications are required to guarantee efficient cooperation among UAVs and timely coordination with the rescue team. Specifically, the uplink video streaming, from UAV fleet to the BS, and device-to-device (D2D) communications among UAVs must be guaranteed to enable accurate and optimal civilians tracking. By defining the communication goal as optimal civilian tracking, semantic information is extracted from  video streams and C\&C signals by jointly considering both the age-of-information (AoI) and the value-of-information (VoI).

\begin{figure*}[!ht]
\centerline{\includegraphics[scale=0.12]{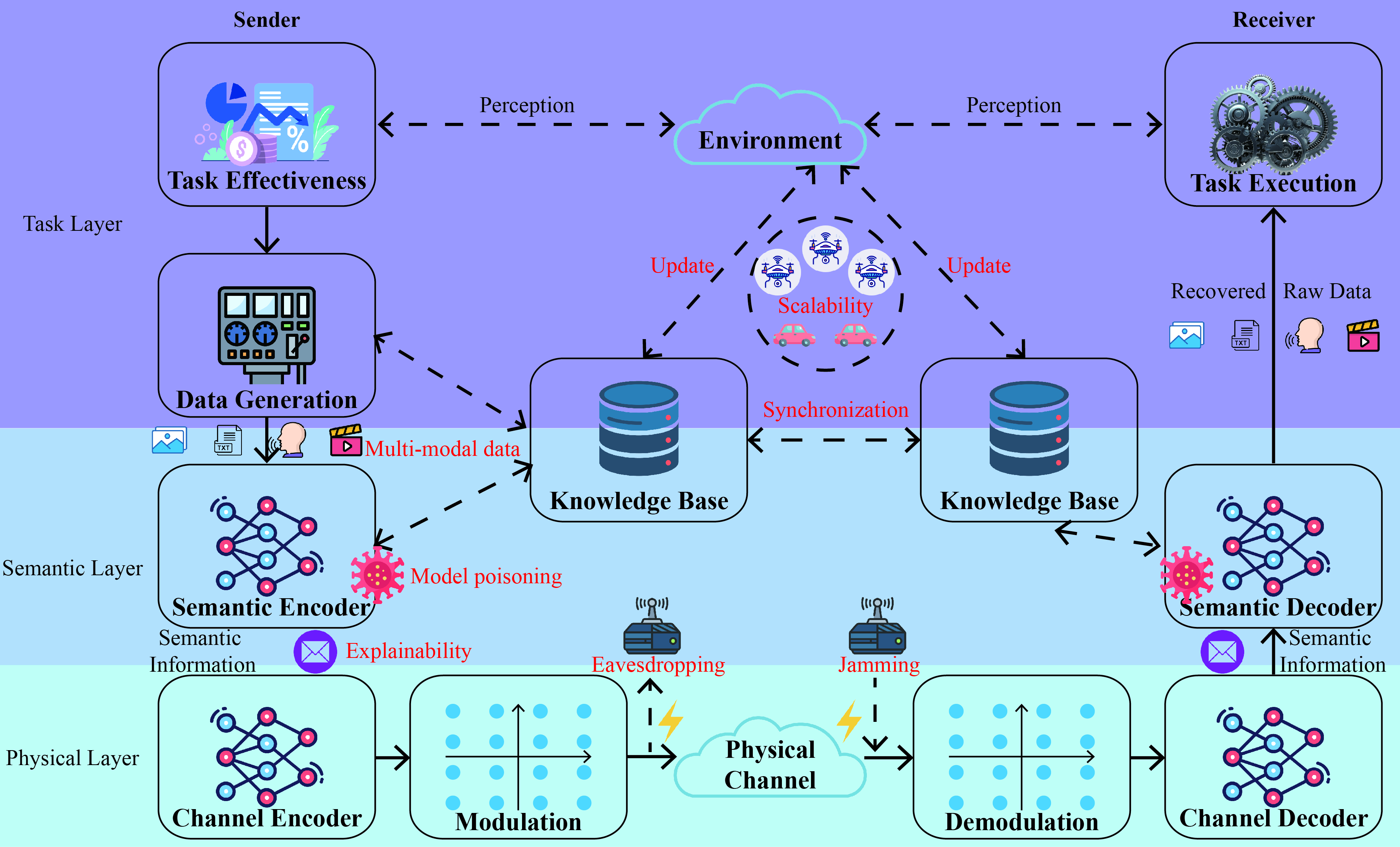}}
\caption{Challenges of vehicle-based services under generic SemCom framework.}
\label{SemCom_framework}
\end{figure*}

\subsubsection{Autonomous Driving}
A quintessential IoV-based service is providing real-time decision support for autonomous vehicles (AVs), especially when they encounter dynamic and unpredictable road conditions \cite{Xu2023}. For example, an AV must respond quickly and safely to sudden roadblocks caused by accidents or ongoing road maintenance. In such safety-critical scenarios, traditional approaches that transmit high-volume raw sensor data (e.g., video or LiDAR point clouds) between vehicles or toward the infrastructure are hampered by excessive latency and bandwidth consumption. SemCom addresses this challenge by enabling an efficient vehicle-centric service model. In this paradigm, each AV uses its onboard sensors to perceive the environment and transforms the raw data into situation-awareness information. This lightweight semantic representation is then transmitted to a traffic monitoring module to derive the appropriate driving decisions, such as initiating a lane change or maintaining the current trajectory. The monitoring module can be deployed in a distributed manner across vehicles or  at the network edge, ensuring timely and reliable coordination.

\section{ Characteristics and Challenges of SemCom for vehicle-based mission-critical services}
In this section, we will analyze the unique characteristics of VbMC services, and investigate the corresponding challenges under SemCom framework as shown in Fig.~\ref{SemCom_framework}.
\subsection{Knowledge Bases Update and Synchronization}
The SemCom framework has introduced the key concept of KBs, which encapsulates the background knowledge related to a specific task, as illustrated in Fig.~\ref{SemCom_framework}. Through KBs synchronization between the transmitter and the receiver, optimal semantic encoders and decoders can be derived for semantic information extraction and reconstruction via end-end training \cite{Wang2025}. However, due to the inherent mobility of vehicles, the wireless network topology changes quickly, leading to a dynamic surrounding environment and fluctuating channel conditions. More importantly,  this dynamic surrounding environment directly influences vehicle-based service provisioning strategies. For example, when a UAV functions as an emergency communication and encounters unexpected weather conditions such as rain or fog, it must adjust its flight strategy according to factors including flight stability, battery capacity, charging stations location, and the accuracy of onboard sensors. Such adaptation is essential to mitigate risks of potential collision or battery depletion. Therefore, there is a pressing need to design adaptive KBs update and synchronization policies, which form the foundation for fine-tuning semantic encoders and decoders in response to instantaneous environmental changes and evolving services provisioning requirements.

\subsection{ Coupled Multi-modal Data}

As shown in the semantic layer of Fig.~\ref{SemCom_framework}, vehicle-based services inherently consist of heterogeneous data types, including C\&C signals and service-related data \cite{Guo2024,Xie2025}. On the one hand, these diverse data types impose heterogeneous Quality of Service (QoS) requirements, e.g., C\&C signals demand high reliability and low latency, while video streaming requires high throughput. On the other hand, C\&C signals and service-related data are closely coupled. For example, in SAR service, each UAV can share instantaneous local C\&C signal and captured video with surrounding UAVs to improve path planning coordination. Notably, the instantaneous C\&C signal is derived not only from locally captured video but also from information exchanged in the previous time slot, such as C\&C signals and captured environmental video from surrounding UAVs. Consequently, instead of transmitting both C\&C signals and captured video independently, the semantic encoder should efficiently extract their coupled features for transmission by understanding the intrinsic relationship between control decisions and video content.

\subsection{SemCom Framework Scalability}

As shown in the task layer of Fig.~\ref{SemCom_framework}, in reality, a swarm of UAVs or a fleet of IoVs must be deployed to accomplish complex tasks jointly via robust coordination \cite{Xu2023,Jiaqi2024}. For example, in autonomous driving, IoV fleets are required to periodically exchange information to improve road safety and traffic efficiency, including the driving strategies, vehicle status and environmental conditions. However, traditional centralized coordination lacks the required flexibility to handle large-scale cooperation and the dynamic topologies in vehicle-based services. To facilitate cooperation, D2D communication combined with multi-hop techniques has emerged as a promising solution to extend coverage and improve scalability among vehicles. Therefore, SemCom must support a hierarchical semantic information encoders and decoders that adapt to the instantaneous VoI at each hop, thereby ensuring efficient and context-aware communication.

\subsection{Semantic Information Explainability}
Although the SemCom framework reduces the required bandwidth for transmission by extracting semantic information only and discarding redundant information from the raw data, the semantic layer of Fig.~\ref{SemCom_framework} reveals a critical limitation of this framework: the lack of explainability. Indeed, abstract semantic representations are often difficult for humans to interpret directly,  leading to a trustworthiness issue. Vehicle-based services, however, require explainable transmitted information to guarantee both reliability and precision in decision making. Existing works have explored explainable SemCom for traditional static text and image transmission, leveraging techniques like text
extraction and segmentation mapping \cite{Wangxijun2025}. However, ensuring the explainability of semantic information in dynamic, safety-critical vehicle-based decision-making services remains an important and unresolved challenge.

\subsection{Threats}
Vehicle-based services are inherently safety-critical, where communication failures can lead to collisions with catastrophic consequences, including loss of human life.  In particular, the degraded communication performance directly impacts IoV services by reducing road safety and traffic efficiency. As shown in the physical and semantic layers of Fig.~\ref{SemCom_framework}, transmitting semantic information instead of raw data increases the likelihood of vulnerability to jamming and eavesdropping attacks. More critically, the use of AI-driven semantic encoders and decoders expands attack surface, introducing novel security risks that go beyond conventional communication threats.

\textbf{ Eavesdropping and Jamming:} 
In traditional bit-oriented communication systems, an eavesdropping attack involves intercepting transmitted data to expose sensitive information, while a jamming attack disrupts ongoing communications by emitting interference or high-power signals, causing bit-level errors or degraded communication quality. In contrast, within the SemCom framework, a semantic eavesdropping attack aims to intercept transmitted signals to infer their semantic meanings and tasks. Similarly, a semantic jamming attack targets the integrity of semantic data, degrading the consistency and quality of transmitted content, thereby preventing reliable task execution at the receiver. Apart from that, the sparse nature of semantic features transmission increases the likelihood of vulnerability to both eavesdropping and jamming attacks, because the transmission characteristics can be easily identified and exploited by adversaries.

\textbf{Semantic Model Poisoning:}
In model poisoning attacks, adversaries manipulate the model parameters to degrade system performance and create biased representations that favor specific inputs, thereby exploiting the semantic encoder and decoder. Such attacks aim to reduce the model's predictive or generative accuracy by manipulating gradient directions or directly modifying weights. The attacker's objective is to mislead the model to misclassifying inputs or inducing uncertainty. In backdoor attacks, adversaries embed hidden triggers in the model at any stage of training (i.e., pre-training, fine-tuning, or alignment). These triggers can be activated only under specific conditions, enabling adversaries to control the semantic encoding and decoding process.

\section{Proposed SemCom Framework for vehicle-based mission-critical services}
\begin{figure*}
    \centering
    \includegraphics[width=1\linewidth]{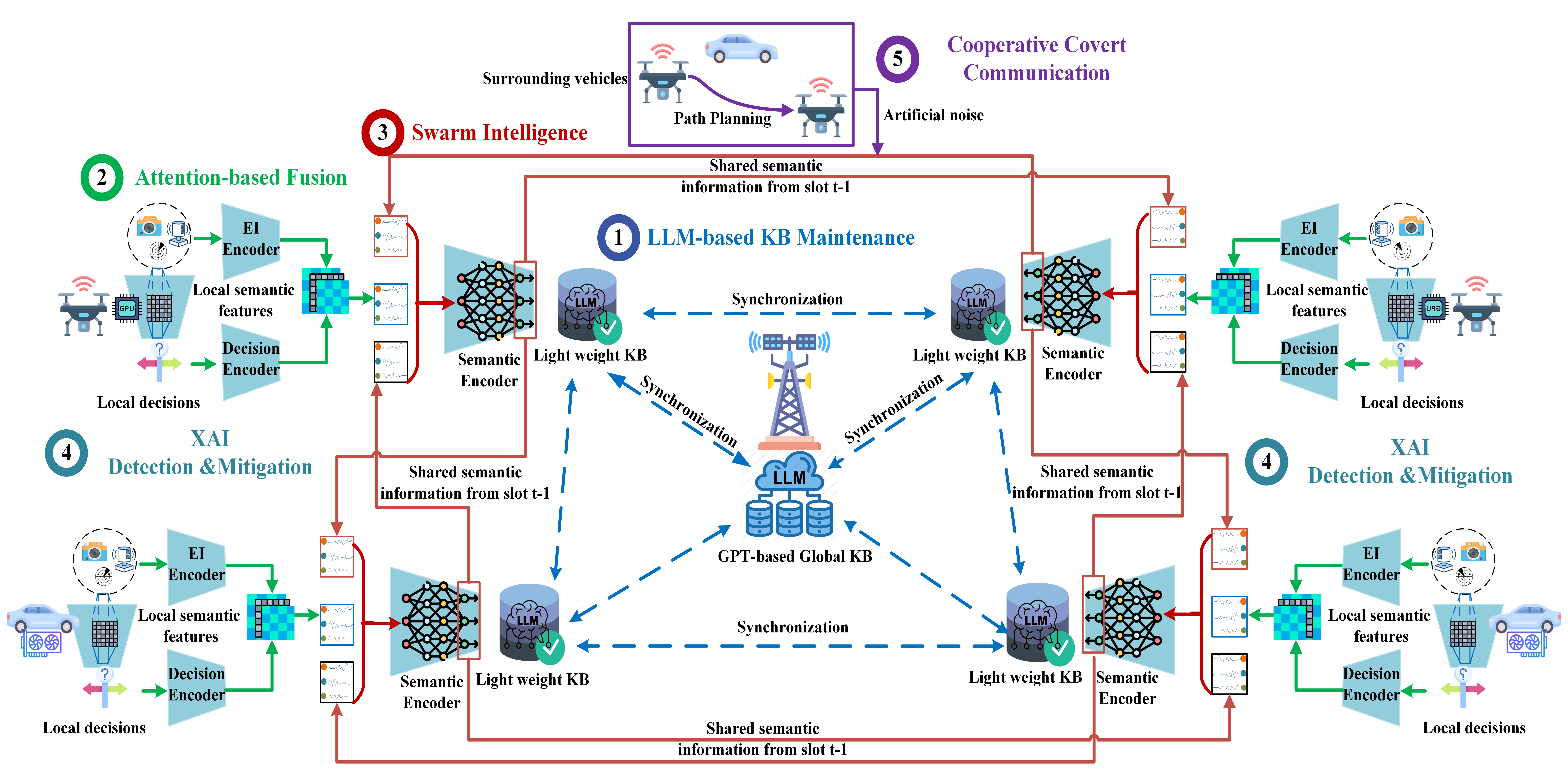}
    \caption{Swarm Intelligence Enabled LLM-based Semantic Communication Framework}
\label{proposed_LLM_framework}
\end{figure*}

In this section, we leverage swarm intelligence combined with LLM to design a novel  SemCom framework tailored for  VbMC services, as shown in Fig. \ref{proposed_LLM_framework}. This framework aims  to solve the above-mentioned challenges through key techniques  discussed below. % as summarized in \ref{solutions}.

\subsection{LLM-based Knowledge Base Maintenance}
The KB is a structured, memory-capable knowledge model that provides relevant semantic knowledge descriptions for raw data. As discussed above, each user is assumed to maintain a local KB; however, significant challenges remain in updating KBs within a dynamic environments and ensuring effective synchronization of the KBs between transmitter and receiver. Specifically, as illustrated in Fig. \ref{proposed_LLM_framework}, a decentralized LLM-based KBs maintenance mechanism for VbMC services is proposed. It leverages the generalization capability of LLM, enabling each vehicle to understand and adapt to the dynamic environments via its LLM-based KB \cite{jiang2024large}. More importantly, considering the limited communication, computation, and storage capabilities of vehicles, lightweight LLM-based KBs with varying model architectures (i.e., different numbers of neural layers) are deployed at the vehicles according to their local capabilities. To ensure effective updating and synchronization of these lightweight LLM-based KBs, a global KB synchronization method is introduced, leveraging a mutual knowledge distillation mechanism. In this method, the global KB (e.g., GPT-5 or Gemini-2.5) can be continuously refined through contributions from individual vehicle's KB using techniques like federated learning and federated distillation \cite{lu2024efficient}.

\subsection{Attention-based Multi-modal Semantic Information Fusion}
As discussed above, each vehicle learns from the collected service-related data, specifically multi-modal environmental information (EI), to derive its local C\&C decisions. Therefore, local C\&C decisions and multi-modal EI are inherently coupled. To facilitate effective cooperation among vehicles, each vehicle may share either its EI or local C\&C decisions with surrounding vehicles, depending on the channel quality and the computing capabilities of these vehicles. Sharing raw multi-modal EI causes a substantial transmission burden and demands high computing capability at the receiving vehicle. To address this challenge, an attention-based multi-modal semantic learning mechanism is proposed, as shown in Fig. \ref{proposed_LLM_framework}, to efficiently process these closely coupled data before sharing. By introducing the attention-based learning, the intrinsic relationships between  local C\&C decisions and multi-modal EI can be effectively learned. Specifically, an EI encoder extracts features from multi-modal environmental data, while a decision encoder extracts the features from the derived local decision-making information.  Then, two potential attention-based learning mechanisms could be employed to further learn the corresponding relationships of the two features: 1) self-attention: the two feature types are concatenated and fed into a self Multi-Head Attention (MHA) module to generate the local semantic features; and 2) cross-attention: the two feature types are first aligned using contrastive learning (e.g., Contrastive Language-Image Pre-training), after which one feature type (e.g., environmental features) is selected as the dominant modality and used as query input to the cross MHA module, while the other feature type (e.g., decision-making features) serves as the source modality, providing Key and Value inputs.

\subsection{Swarm Intelligence for Decentralized Semantic Encoder Update}

To address the scalability challenge discussed in Section III, a swarm intelligence mechanism is incorporated into the proposed distributed semantic framework, as shown in Fig. \ref{proposed_LLM_framework}. Within this framework, each vehicle can receive semantic information shared by other vehicles to enrich its perception of the entire environment, and assist in relaying semantic information to the broader network via D2D communications. This ensures both cooperative awareness and scalable information dissemination. Specifically, in time slot $n$, each vehicle receives semantic information from surrounding vehicles generated in time slot $t-1$. Then, the received semantic information of time slot $t-1$, together with the vehicle's local semantic information of time slot $t$ are fed into the semantic encoder to generate richer, context-aware representations. These enhanced semantic features are then shared with surrounding vehicles. The sharing mechanism can be modeled using swarm intelligence, with the objective of achieving comprehensive real-time understanding of the overall task environment. This process further emphasises  optimizing the acquisition  of  context-aware information with minimum exchange of semantic information, especially in heterogeneous network scenarios.

\subsection{XAI-Guided Semantic Feature Filter}
\begin{figure}[!ht]
\centering
\subfloat[XAI-guided semantic feature extraction for building interpretable semantic encoders in the SemCom framework.]
{\includegraphics[scale=0.32]{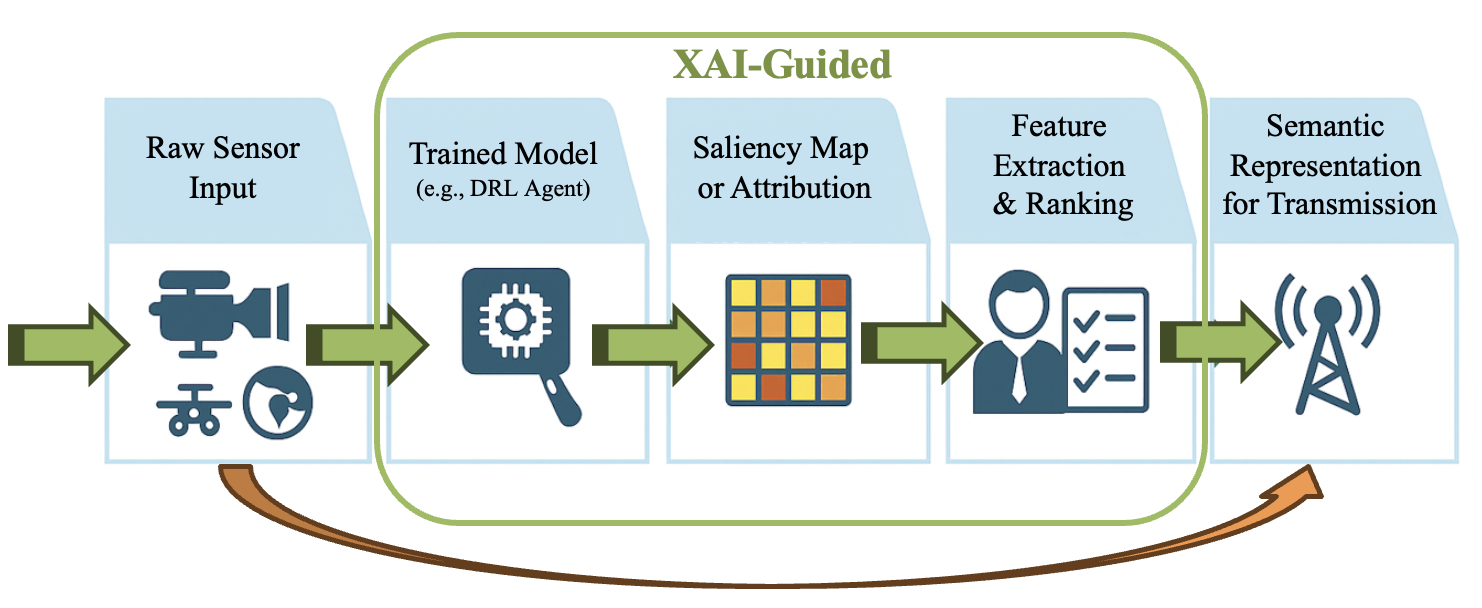}} \\
\centering
\subfloat[Model poisoning detection (left) and mitigation solutions (right)]
{\includegraphics[scale=0.145]{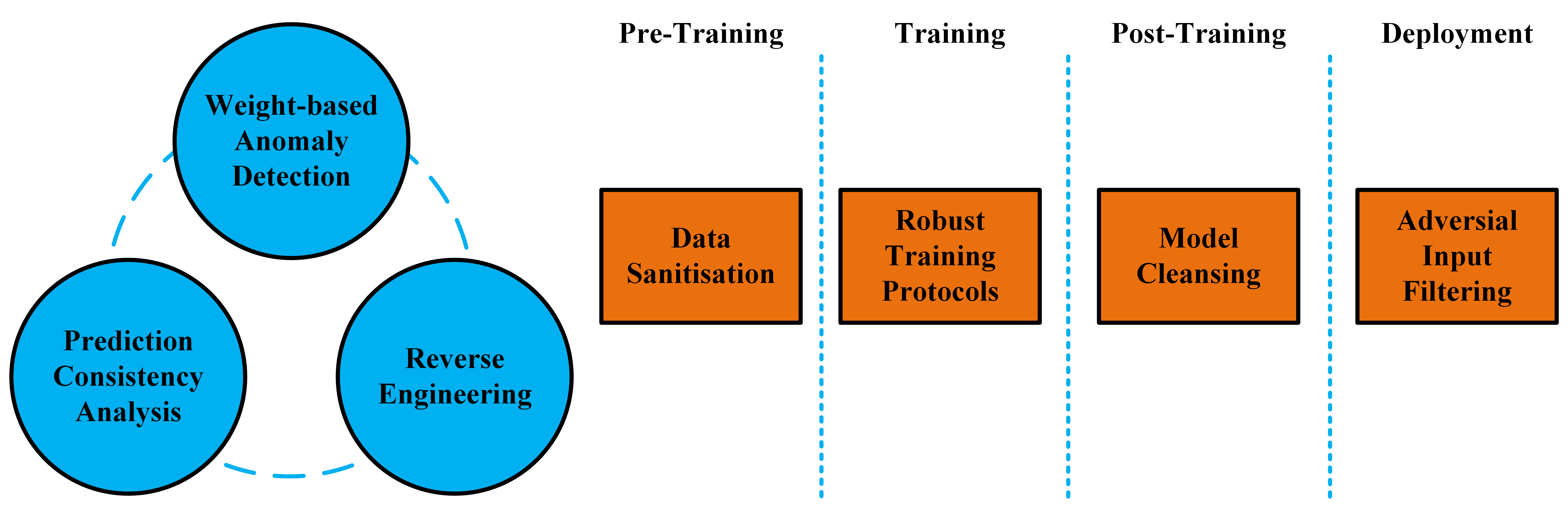}}
\caption{Proposed explainability and security solutions.}
\label{fig:solutions}
\end{figure}

Within the SemCom framework for VbMC services, transmitted information must be both compact and interpretable, not only to improve user trust but also to support reliable and timely decision-making. To achieve this, XAI techniques are utilized to design an explainable semantic encoder that efficiently extracts the critical semantic features that influence the model's decisions, thereby reducing data transmission loads without sacrificing decision accuracy. The proposed XAI-guided semantic filtering mechanism consists of three key components, as shown in Fig.\ref{fig:solutions} (a). First, given a trained decision-making model, typically a Deep Reinforcement Learning (DRL) algorithm used in VbMC services, XAI techniques, such as SHapley Additive exPlanations (SHAP) or Gradient-weighted Class Activation Mapping (Grad-CAM), are applied to identify high-impact features from structured input or salient regions within image data. Second, the generated explanations are analyzed to rank the importance of semantic components. This step varies depending on the VbMC services context, e.g., congestion relief or UAV navigation. A subset of top-ranked features is then selected, either automatically or with human-in-the-loop guidance. Finally, these selected features are encoded into compact semantic representations and transmitted to downstream modules, replacing large-scale raw data. In this way, the proposed XAI-guided semantic feature filter supports various XAI methods for both structured and image data, enabling explainable semantic encoding that enhances transmission efficiency and trust in VbMC services.

\subsection{Cooperative Covert Communication}
Covert communication aims to achieve  undetectable transmissions by embedding the transmission
signals into environmental or artificial noise. Its effectiveness is usually evaluated using metrics such as covert communication capacity, detection error probability (DEP), and covert rate. To overcome the aforementioned security challenges, including eavesdropping and jamming attacks, a cooperative covert communication scheme for the VbMC services is proposed, as illustrated in Fig.~\ref{proposed_LLM_framework}. In this scheme, the inherent dynamics of vehicle positions and cooperative behaviors are utilized to introduce randomness, thereby reducing the likelihood of detection by a warden. Specifically, transmitter vehicle trajectory optimization problem can be formulated to maximize VbMC service metrics while satisfying covert communication constraints. Furthermore, for highly sensitive information, the transmitter can request assistance from surrounding idle vehicles to generate artificial noise, providing an additional layer of concealment for the transmission.

% \subsection{Xiaolan and Marios}
\subsection{Poisoning Detection and Mitigation}

Addressing poisoning attacks against SemCom is a complex challenge, that can be classified into detection and mitigation as shown in Fig.\ref{fig:solutions} (b).

\subsubsection{Detection}

The proposed detection solution integrates three complementary mechanisms: weight-based anomaly detection, prediction consistency analysis inspired by graph variance methods, and efficient reverse engineering with adaptive thresholding that automatically calibrates detection sensitivity based on network architecture and application domain. Together, these mechanisms allow fast detection and maintain high true positive rates across diverse attack vectors, including emerging attacks on diffusion models and graph neural networks. Apart from that, when combined with XAI techniques, this solution can provide transparent justifications for detection outcomes, and support automated forensic analysis in identifying complex  cross-modal backdoor attacks.

\subsubsection{Mitigation}
The proposed mitigation solution addresses vulnerabilities across all stages of ML lifecycle, including pre-training, training, post-training, and deployment phases simultaneously.  In the pre-training phase, data sanitisation is performed using poisoned sample detection, leveraging  recent advances in activation clustering and gradient analysis to filter malicious inputs. More importantly, robust training protocols, such as differential privacy, are utilized to provide provable security bounds. In post-training phase, it is essential to preserve model integrity by eliminating  backdoors or poisonous parameters via selective neuron pruning, guided by activation alignment principles. Finally, when the model is deployed within  the SemCom framework, real-time input validation is integrated to detect adversarial triggers across multiple modalities, ensuring secure and trustworthy operation in dynamic environments.

\section{Case Study}
To demonstrate the effectiveness of the proposed SemCom framework, we present a case study focusing on TLC optimization using the proposed XAI-guided semantic feature extraction. In this setup, UAVs equipped with cameras capture real-time traffic conditions and transmit the data stream to the TLCs. The DRL-based TLC optimization algorithm is trained using full image inputs and operates at every simulation timestep, dynamically deciding whether to switch to the next signal phase or extend the current one. The reward function is designed to minimize queue lengths and vehicle waiting times while penalizing excessive or frequent signal changes to guarantee the desired efficiency.

\begin{figure}[!ht]
    \centering
    \includegraphics[width=1\linewidth]{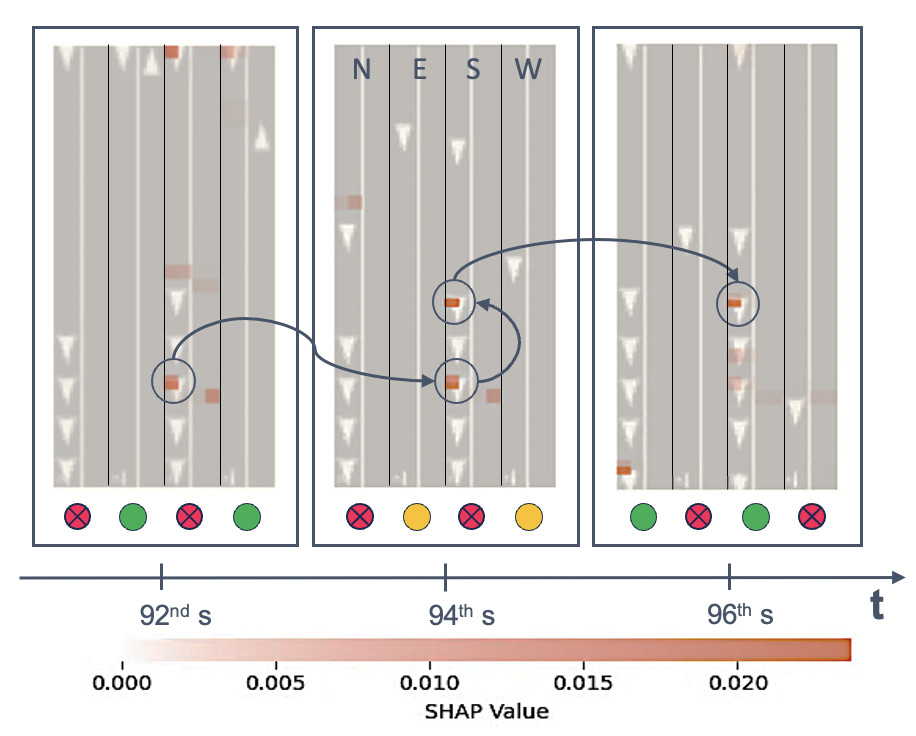}
    \caption{SHAP saliency maps during one full traffic signal phase switching. }
\label{xai_shap}
\end{figure}

To improve interpretability and reduce communication costs, SHAP saliency maps are generated from camera snapshots to interpret the decisions of the pre-trained DRL model. SHAP values quantify the change in model output when the decomposed image regions are occluded, thereby presenting feature importance. In this case study, through human-in-the-loop analysis, a key semantic feature is identified: the location of the last vehicle in the first platoon on each incoming road, as shown in Fig.\ref{xai_shap}. Large SHAP values highlight areas in the traffic image that exert a significant influence on the DRL agent's actions. As time advances (from left to right), vehicles accumulate on the North-South (NS) roads, forming longer platoons that indicate the  need for phase switching. The largest SHAP value highlights the location of the last vehicle in the first platoon on the S-incoming road, emphasising its critical role in the model’s decision making. This semantic feature consistently exerts a strong influence on the DRL model’s switching behavior, enabling efficient coordination of  intersection traffic, thus guiding the subsequent explainable semantic encoding process.

\begin{table}[!ht]
\caption{Performance  evaluation results comparison: Static TLC vs DRL-based TLC vs XAI-DRL-based TLC}
\begin{center}
	\begin{tabular}{|c|c|c|c|c|}
\hline
\textbf{Method} & \textbf{Travel Time} & \textbf{Delay} & \textbf{Training time} & \textbf{Comm. cost}\\
\hline
Static & 32.95s & 20.68s & - & - \\
\hline
DRL & 23.30s & 11.06s & 24.29s & 24580B \\
\hline
XAI-DRL & 21.09s & 8.82s & 11.65s & 20B \\
\hline
\end{tabular}
\label{tab:simple_results}
\end{center}
\end{table}

The XAI-guided semantic vector encodes traffic information using  four scalar values (one per road) representing vehicle positions, along with the current traffic light phase index. Performance evaluation results highlight that raw camera images (e.g., 128×64×3 bytes) combined with one signal phase index require approximately  24,580 bytes per input data. In contrast, the proposed XAI-guided semantic information extraction scheme transmits only the location of the last-vehicle in the first platoon (4 values) plus the phase index, resulting in a compact representation of just 20 bytes at a four-way intersection. This approach reduces communication overhead by more than 90\% and decreases model training time by approximately 21\% while maintaining effectiveness in congestion mitigation, as summarised in Table \ref{tab:simple_results}.

\section{Conclusion}
In this article, we propose a swarm intelligence-enabled and LLM-based SemCom framework tailored for Vehicle-based Mission-Critical (VbMC) services. We begin by analyzing typical vehicle-based information-critical and infrastructure-critical  services within the context of the generic SemCom framework, identifying key data types, link types, and semantic information. We then highlight the unique characteristics and challenges associated with the application of this generic SemCom framework to VbMC services, including issues related to synchronization, heterogeneity, cooperation, explainability, and security. To address these challenges,  we propose an advanced SemCom framework that integrates swarm intelligence and LLM, leveraging cutting-edge AI and communication technologies. A case study focused on UAVs for rapid congestion relief demonstrates the practicality of the proposed framework.  Using an XAI-guided semantic extraction scheme, the case study verifies the framework’s effectiveness in reducing transmitted data volume while guaranteeing the necessary level of explainability.
\bibliographystyle{IEEEtran}
\bibliography{IEEEabrv,Main}

\vskip -2.5\baselineskip plus -1fil
\begin{IEEEbiographynophoto}{Hui Zhou} is currently an Assistant Professor at Coventry University. His research interests include NTN, and semantic communication.
\end{IEEEbiographynophoto}

\vskip -3\baselineskip plus -1fil
\begin{IEEEbiographynophoto}{Jiaying Guo} is currently a Post-doc at University College Dublin.  Her research interests include the multi-agent systems of intelligent transportation.
\end{IEEEbiographynophoto}

\vskip -3\baselineskip plus -1fil
\begin{IEEEbiographynophoto}{Marios Aristodemou} is currently a Post-doc at the University of York. His research interests include security and trustworthiness, personalized distributed learning, privacy preserving, and lifelong learning.

\end{IEEEbiographynophoto}

\vskip -3\baselineskip plus -1fil
\begin{IEEEbiographynophoto}{Zhaoyang Du} is currently a Post-doc at the University of Electro-Communications.  His current research
interests include delay tolerant networks, vehicular
ad hoc networks, and IoT.

\end{IEEEbiographynophoto}

\vskip -3\baselineskip plus -1fil
\begin{IEEEbiographynophoto}{Shen Wang} is currently an Assistant Professor at University College Dublin. His research interests include connected autonomous
vehicles, explainable artificial intelligence, and security and privacy for
mobile networks.

\end{IEEEbiographynophoto}

\vskip -3\baselineskip plus -1fil
\begin{IEEEbiographynophoto}{Xiaolan Liu} is currently an Assistant Professor at the University of Bristol. Her current
research interests include wireless distributed learning, multiagent RL for edge
computing, and ML for wireless
communication optimization.
\end{IEEEbiographynophoto}

\vskip -3\baselineskip plus -1fil
\begin{IEEEbiographynophoto}{Soufiene Djahel} is currently a Professor at Coventry University. His research interests include the design and evaluation of communication, planning, optimization and security algorithms. 
\end{IEEEbiographynophoto}

\vskip -3\baselineskip plus -1fil
\begin{IEEEbiographynophoto}{Celimuge Wu} is currently a Professor at the University of Electro-Communications. His research interests
include Vehicular Networks, Edge Computing, IoT,
and AI for Wireless Networking and Computing
\end{IEEEbiographynophoto}

\end{document}